\documentclass[%
superscriptaddress,
 amsmath,amssymb,
 aps,
prb,
twocolumn,
floatfix,
]{revtex4-2}
\usepackage{mhchem}
\usepackage{gensymb}
\usepackage{booktabs}
\usepackage{graphicx}
\usepackage{dcolumn}
\usepackage{bm}
\usepackage[dvipsnames]{xcolor}
\usepackage{float}

\begin{document}
\preprint{APS/123-QED}

\title{Electronic and structural properties of RbCeX$_2$ (X$_2$: O$_2$, S$_2$, SeS, Se$_2$, TeSe, Te$_2$)}%

\author{Brenden R. Ortiz$^{\dagger}$}
 \email{ortiz.brendenr@gmail.com}
 \affiliation{Materials Department and California Nanosystems Institute, University of California Santa Barbara, Santa Barbara, CA, 93106, United States}%
 
\author{Mitchell M. Bordelon}
 \affiliation{Materials Department and California Nanosystems Institute, University of California Santa Barbara, Santa Barbara, CA, 93106, United States}%
 
 \author{Pritam Bhattacharyya}
 \affiliation{Institute for Theoretical Solid State Physics, Leibniz IFW Dresden, Helmholtzstr.~20, 01069 Dresden, Germany}%
 
 \author{Ganesh Pokharel}
 \affiliation{Materials Department and California Nanosystems Institute, University of California Santa Barbara, Santa Barbara, CA, 93106, United States}%
 
 \author{Paul M. Sarte}
 \affiliation{Materials Department and California Nanosystems Institute, University of California Santa Barbara, Santa Barbara, CA, 93106, United States}%
 
 \author{Lorenzo Posthuma}
 \affiliation{Materials Department and California Nanosystems Institute, University of California Santa Barbara, Santa Barbara, CA, 93106, United States}%

 \author{Thorben Petersen}
 \affiliation{Institute for Theoretical Solid State Physics, Leibniz IFW Dresden, Helmholtzstr.~20, 01069 Dresden, Germany}%

 \author{Mohamed S. Eldeeb}
 \affiliation{Institute for Theoretical Solid State Physics, Leibniz IFW Dresden, Helmholtzstr.~20, 01069 Dresden, Germany}%

\author{Garrett E. Granroth}
 \affiliation{Neutron Scattering Division, Oak Ridge National Laboratory, Oakridge, TN, 37830, United States}%

 \author{Clarina R. Dela Cruz}
 \affiliation{Neutron Scattering Division, Oak Ridge National Laboratory, Oakridge, TN, 37830, United States}%

\author{Stuart Calder}
 \affiliation{Neutron Scattering Division, Oak Ridge National Laboratory, Oakridge, TN, 37830, United States}%

\author{Douglas L. Abernathy}
 \affiliation{Neutron Scattering Division, Oak Ridge National Laboratory, Oakridge, TN, 37830, United States}%
 
 \author{Liviu Hozoi}
 \affiliation{Institute for Theoretical Solid State Physics, Leibniz IFW Dresden, Helmholtzstr.~20, 01069 Dresden, Germany}%

\author{Stephen D. Wilson}
 \email{stephendwilson@ucsb.edu}
 \affiliation{Materials Department and California Nanosystems Institute, University of California Santa Barbara, Santa Barbara, CA, 93106, United States}%

\date{\today}

\begin{abstract}
Triangular lattice delafossite compounds built from magnetic lanthanide ions are a topic of recent interest due to their frustrated magnetism and realization of quantum disordered magnetic ground states. Here we report the evolution of the structure and electronic ground states of RbCe$X_2$ compounds, built from a triangular lattice of Ce$^{3+}$ ions, upon varying their anion character ($X_2$= O$_2$, S$_2$, SeS, Se$_2$, TeSe, Te$_2$).  This includes the discovery of a new member of this series, RbCeO$_2$, that potentially realizes a quantum disordered ground state analogous to NaYbO$_2$.  Magnetization and susceptibility measurements reveal that all compounds manifest mean-field antiferromagnetic interactions and, with the exception of the oxide, possess signatures of magnetic correlations onset below 1 K. The crystalline electric field level scheme is explored via neutron scattering and \textit{ab initio} calculations in order to model the intramultiplet splitting of the $J=5/2$ multiplet.   In addition to the two excited doublets expected within the $J=5/2$ manifold, we observe one extra, local mode present across the sample series.  This added mode shifts downward in energy with increasing anion mass and decreasing crystal field strength, suggesting a long-lived anomalous mode endemic to anion motion about the Ce$^{3+}$ sites.
\end{abstract}

\maketitle

\section{Introduction}

The search for material systems that play host to quantum spin liquid (QSL) states or intrinsic quantum disorder is a key impetus for continued exploration of frustrated motifs and magnetic frustration in new materials. A prototypical spin-liquid candidate is typically envisioned to combine highly quantum magnetic moments ($S=1/2$), a frustrated lattice motif, and tailored anisotropies that favor persistent quantum fluctuations \cite{inttri_anderson1973resonating, inttri_anderson1987resonating, inttri_balents2010spin, inttri_broholm2020quantum, inttri_lee2008end, inttri_savary2016quantum, inttri_witczak2014correlated,inttri_li2020spin}.  The triangular lattice is a canonical manifestation of these ingredients -- one whose ground state depends critically on the character of the nearest-neighbor interaction, the size of the individual moments, and the types and strengths of anisotropies. 

The design and realization of materials that unambiguously support QSL phenomenon remains challenging. Within triangular lattices, several prospective organic materials (e.g. $\kappa$-(BEDT-TTF)$_2$Cu$_2$(CN)$_3$, EtMe$_3$Sb[Pd(dmit)$_2$]$_2$) \cite{tri_itou2008quantum,tri_lee2005u} have been proposed, and the discovery of new candidate materials is a persistent theme in solid state chemistry. A recent frontier in inorganic materials has appeared in new triangular lattice materials built from rare-earth ions that exhibit ground state doublets and $S_\text{eff} = 1/2$ moments. Single-hole $f^{13}$ Yb$^{3+}$ and single-electron $f^1$ Ce$^{3+}$ ions in low-symmetry ligand fields are excellent candidates for stabilizing $S_\text{eff} = 1/2$ states with strong fluctuations, and both are well-known constituents of the triangular lattice systems. Yb-containing oxide compounds like NaYbO$_2$ \cite{nyo_bordelon2019field} and YbMgGaO$_4$ \cite{yb_kimchi2018valence,yb_li2017nearest} have recently been shown to exhibit quantum disordered ground states consistent with a QSL.

Promising inorganic structures for hosting QSL phenomenon among Ce$^{3+}$ and Yb$^{3+}$ moments require magnetic cations located at high-symmetry Wyckoff sites \textit{and} an innate resistance to anti-site and defect formation. Compounds crystallizing in the form $A$$R$$X_2$ (where $A$ is an alkali ion, $R$ is rare-earth ion, and $X$ is a chalcogen anion) are excellent candidates and commonly crystallize in the $R\overline{3}m$ $\alpha-$NaFeO$_2$ structure, which possesses a structurally perfect triangular lattice of rare-earth ions. The broader $A$$R$$X_2$ family is extremely diverse \cite{bordelon2021frustrated,112gen_cantwell2011crystal,112gen_clos1970deux,112gen_dong2008structure,112gen_fabry2014structure,112gen_hashimoto2002structures,xing2019crystal} and when decorated with $S_\text{eff}=1/2$ moments, a large degree of magnetic frustration arises \cite{112fr_baenitz2018naybs,112fr_hashimoto2003magnetic,112fr_liu2018rare} with ground states that vary with the character of the surrounding anions. 

Systematic studies of ligand field effects and perturbations of the anionic sublattice allow for an important window into the physics of $ARX_2$ compounds. Perturbations of the anionic sublattice allow for the investigation of the influence of local polyhedral distortions, anionic disorder, and sterics (e.g. bond angle/distance) on the evolution of the magnetic ground state. Alloying the anion sublattice also allows for strong chemical disorder on the $X$-site while preserving the anisotropies and orbital character of triangular network of $R$-site ions. As one example, in single-hole $R$=Yb $A$$R$$X_2$ systems, oxide variants such as NaYbO$_2$ exhibit an unconventional quantum disordered ground state \cite{nyo_bordelon2019field,nyo_ding2019gapless,nyo_ybranjith2019field}. In contrast, their selenide counterparts exhibit nearly static, quasi-two dimensional correlations \cite{scheie2021witnessing,PhysRevB.100.220407}. While the single-electron counterparts KCeO$_2$ and KCeS$_2$ both develop signs of magnetic ordering at low temperature \cite{kco_bordelon2021magnetic, kulbakov2021stripe}, a systematic study exploring ligand field effects for Ce$^{3+}$ ions in $A$$R$$X_2$ remains absent.

In this work, we present a comprehensive study of the RbCeX$_2$ delafossite compounds as a function of the chalcogen site. We demonstrate that a full solid solution exists between the RbCeTe$_2$--RbCeSe$_2$--RbCeS$_2$, allowing us to tune the local chemistry while preserving the magnetic Ce$^{3+}$ triangular lattice. Our work also presents the discovery of a new oxide delafossite, RbCeO$_2$. Neutron and synchrotron x-ray diffraction data characterize the structural evolution, including local distortions of the ligand ions, and susceptibility/magnetization measurements are used to correlate the corresponding changes in magnetic properties.  Inelastic neutron scattering measurements combined with \textit{ab initio} modeling further examine the evolution of the crystalline electric field ground state and intramultiplet splitting of the Ce$^{3+}$ $J=5/2$ spin-orbital manifold.  Of particular interest, our combined data establish that RbCeO$_2$ possesses the strongest antiferromagnetic exchange field, hosts the strongest distortion to its local Ce$X_6$ octahedra, and uniquely exhibits no signatures of long-range order or moment freezing. The properties of RbCeO$_2$ resemble those of NaYbO$_2$, and suggest that RbCeO$_2$ may exhibit an analogous disordered ground state. We further demonstrate that the anomalous local electronic mode identified in KCeO$_2$ \cite{kco_bordelon2021magnetic}, proposed to manifest as a long-lived vibronic state, also appears across the entire RbCe$X_2$ series.  The energy of this mode shifts downward with increasing mass of the anion oscillator and with decreasing crystal field strength, further linking this mode to an anomalous coupling between Ce$^{3+}$ spin-orbital states and anion degrees of freedom.

\section{Methods}

\subsection{Synthesis}
Powders of RbCeX$_2$ compounds were produced using Rb metal (Alfa, 99.75\%), Ce rod (Alfa, 99.9\%), S chunk (Alfa, 99.999\%), Se shot (Alfa, 99.999\%), Te chunk (Alfa, 99.999\%), and CeO$_2$ powder (Alfa, 99.99\%). RbCeO$_2$ was synthesized by placing Rb metal and CeO$_2$ at a molar ratio of 1.25:1 into 2\,mL Al$_2$O$_3$ crucibles that were sealed under argon into stainless steel tubes. These tubes were heated to 800\degree C for 48\,h at 200\degree C/h. The tubes are cut open under argon and the RbCeO$_2$ extracted from the crucible afterwards. The excess Rb ensures Rb saturation within the RbCeO$_2$. All non-oxide RbCeX$_2$ compounds are synthesized through mechanochemical methods. The Rb metal, Ce shavings, and the appropriate chalcogen were added in near stoichiometric ratios 1.05:1:2 (Rb:Ce:X) to a tungsten carbide ball mill vial. Samples were milled for 60\,min using a Spex 8000D mill. The resulting powders were placed in 2\,mL Al$_2$O$_3$ crucibles, sealed within fused silica ampoules, and annealed under approximately 0.8\,ATM of argon for 18\,h. The anneal temperatures were tuned based on the chalcogen to ensure high crystallinity throughout the series (Te$_2$: 600\degree C, TeSe: 600\degree C, Se$_2$: 650\degree C, SeS: 650\degree C, S$_2$: 650\degree C).

After heat treatment, each compound exhibits extremely strong coloration which acts as a qualitative measure of sample quality. In their pristine state: RbCeO$_2$ powder is blood-red, RbCeS$_2$ is dark yellow, RbCeSe$_2$ is a dark orange, RbCeTe$_2$ is a metallic purple. The subsequent alloys exhibit colors that are conceptually consistent with the end members: RbCeSeS is a dark yellow-orange, RbCeTeSe is a dark fuchsia. Perhaps owing to the relatively low stability of trivalent Ce$^{3+}$, the RbCeX$_2$ series of compounds are sensitive to air and water. The degree of sensitivity varies with anion, from extremely ($<$1\,s) sensitive in the case of RbCeO$_2$ to ($<$1\,m) in the case of RbCeTe$_2$. All compounds darken and blacken with exposure to air, with singular exception of RbCeO$_2$, which turns white. We note that single crystals of the compounds do not exhibit the same degree of sensitivity, and while this manuscript will focus on the polycrystalline materials, macroscopic single crystals may present a means to circumvent the air sensitivity. 

\subsection{Scattering Measurements}

High-resolution synchrotron x-ray powder diffraction data were collected using beamline 11-BM at the Advanced Photon Source (APS), Argonne National Laboratory using an average wavelength of 0.457900\,\AA. Discrete detectors covering an angular range from -6 to 16\degree\, (2$\rm{\theta}$) are scanned over a 34\degree\, range, with data points collected every 0.001\degree\, and scan speed of 0.01\degree/s. Due to the air sensitivity of the materials, small quantities of each RbCeX$_2$ were diluted with amorphous SiO$_2$ in a glovebox and sealed under argon in flame-tipped amorphous SiO$_2$ capillaries. These capillaries were nested within kapton sleeves and held in place with a small amount of modeling clay. Laboratory powder diffraction utilized a Panalytical Empyrean diffractometer (Cu K$\alpha$, 1.54\AA) in the Bragg-Brentano geometry. 

Inelastic neutron scattering (INS) data were collected on the wide Angular-Range Chopper Spectrometer (ARCS) at the Spallation Neutron Source (SNS) at Oak Ridge National Laboratory (ORNL). Multiple incident energies of $E_i = 300, 150, 60$\,meV  at $T =$ 5 and 300\,K were used to map out the crystalline electric field (CEF) excitations of polycrystalline samples of RbCeX$_2$ (X$_2$: O, S, SeS, Se, TeSe, Te) loaded in aluminum sample canisters.   A Fermi chopper, with slits spaced 1.52 mm apart of a 1.5, was spun at 420, 600, and 600 Hz to select $E_i$'s of 60, 150, and 300 meV, respectively. For $E_i$ =60 meV and 150 meV, the blades of the selected Fermi chopper had a radius of curvature of 0.580 m.  The blades in the Fermi chopper used for $E_i=300$ meV had a radius of curvature of 1.52 m. The background contribution from the aluminum sample canisters was subtracted from the data sets by obtaining an empty canister measurement at each E$_i$ and $T$. Energy cuts of the data were integrated to analyze the CEF excitations by fitting the peaks to a Gaussian function approximating the energy resolution of the instrument.

Neutron powder diffraction was performed at the HB-2A instrument at ORNL at room temperature using 1.54\AA\; neutrons were selected from a vertically focused Ge(115) wafer-stack monochromator. This experiment utilized the same samples as the inelastic experiment on ARCS, and powders were left sealed in their Al cans between measurements. 

Structure solution of the RbCeO$_2$ was performed using charge flipping methods\cite{oszlanyi2008charge,coelho2007charge,coelho2018topas} on the x-ray diffraction data. Further refinement of the initial structure was performed using the neutron data to determine the oxygen sites. Finally, for the best structural fidelity across the series, we performed co-refinement of the neutron and x-ray diffraction data using {\sc topas academic v6} \cite{coelho2018topas}. To account for the aluminum can in the neutron data, a simple Pawley refinement was performed to remove any Al contributions. No corrections were needed for the kapton or silica additives in the synchrotron data due to the high intensity of the diffracted beam. 

\subsection{Magnetization Measurements}

For magnetization measurements, approximately 10-15\,mg of each powder was sealed into a  polypropylene (Formolene 4100N) capsule using a small amount of parafilm to maintain an air-free environment during sample transfer. Temperature-dependent magnetization data between 300\,K and 1.8\,K were collected using a Quantum Design Squid Magnetometer (MPMS3) in vibrating-sample measurement mode (VSM). High-field magnetization measurements up to 14\,T were performed on a Dynacool Physical Property Measurement System (PPMS) equipped with a VSM attachment. 

AC susceptibility results were performed on a Quantum Design Dynacool Physical Property Measurement System (PPMS) equipped with a dilution refrigerator insert and AC susceptibility module. To circumvent the air sensitivity of the RbCeX$_2$ powders, we developed a simple protective method using paraffin wax encapsulation. Small quantities of wax were melted in a glass vial inside an argon glove box. Once the wax was molten, RbCeX$_2$ powders were mixed into the molten wax at an approximate mass ratio of 20:80 wax:RbCeX$_2$. The vial was removed from heat and continually stirred until the mixture was homogeneous and the wax solidified. The mixture was mechanically removed from the vial and loaded into a 3\,mm diameter stainless steel die press and consolidated. The resulting pellet was sectioned into 2\,mm cubes. This preparation was sufficient to maintain material quality (coloration, etc.) for up to 1\,hr. Measurements on wax blanks revealed that the wax contributes a completely temperature \textit{independent} background. Thus, a trivial scaling to the 1.8\,K magnetization data collected \textit{via} VSM eliminates the influence of the wax background.

\subsection{Crystalline Electric Field Analysis}

The analysis of crystalline electric field (CEF) schemes of RbCe$X_2$ ($X_2$: O$_2$, S$_2$, SeS, Se$_2$, TeSe, Te$_2$) followed the procedure performed for KCeO$_2$ \cite{kco_bordelon2021magnetic}. These materials crystallize in the same crystal structure with trivalent Ce$^{3+}$ in a local $D_{3d}$ environment. Trivalent Ce ions contain total angular momentum $J = 5/2$ ($L$ = 3, $S$ = 1/2) that is split into three Kramers doublets by the $D_{3d}$ point group environment. A CEF Hamiltonian with parameters $B_n^m$ and Steven’s operators $\hat{O}_m^n$ in $D_{3d}$ symmetry is known \cite{stevens1952matrix}.

\begin{figure}
\includegraphics[width=\linewidth]{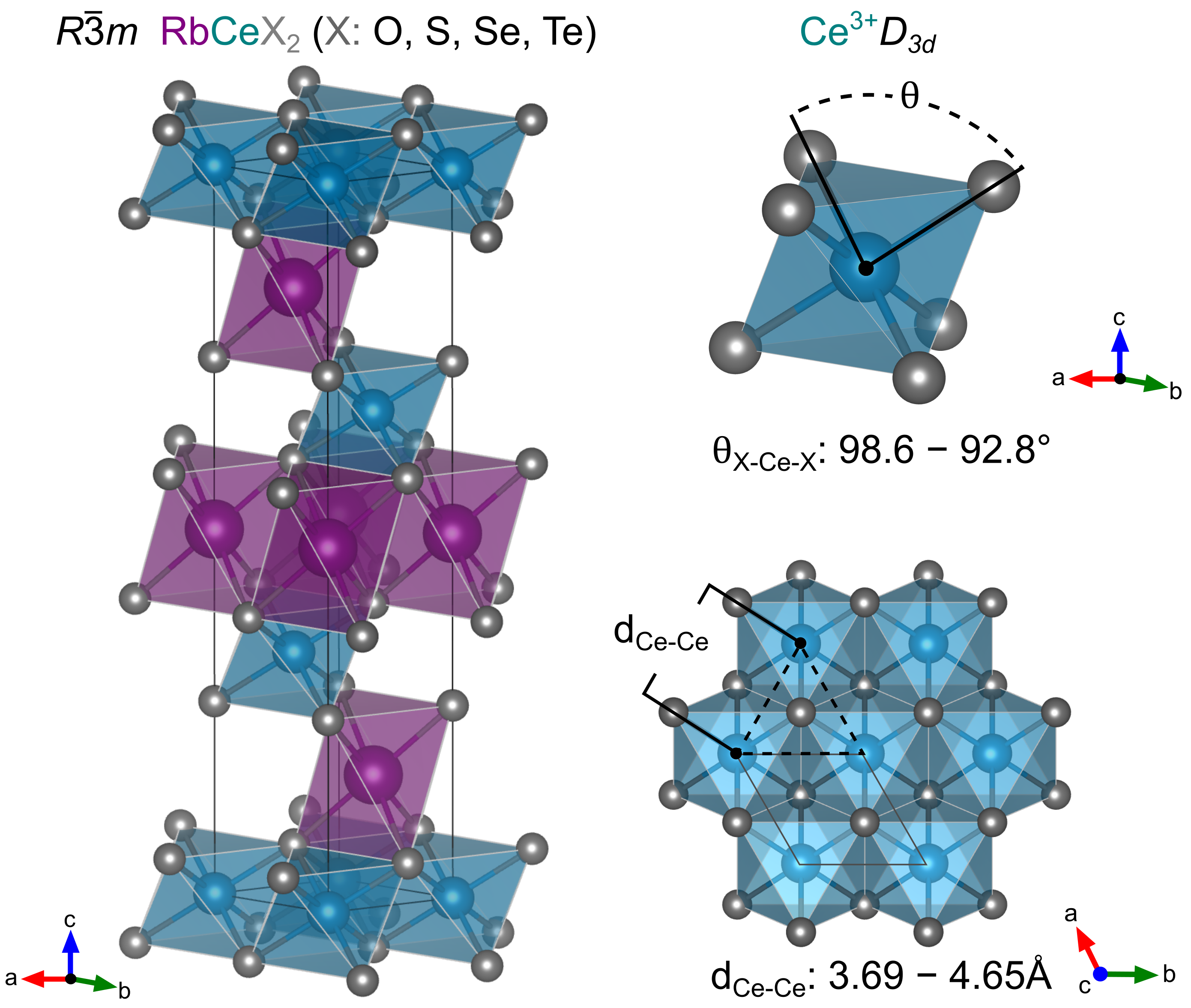}
\caption{The RbCeX$_2$ family of compounds crystallize in the layered $R\overline{3}m$ crystal structure with a triangular lattice of Ce$^{3+}$ ions. The Ce$^{3+}$ exists in a local octahedral $D_{3d}$ environment. The octahedra are trigonally distorted (sheared), as parameterized by the angle $\theta$. The degree of distortion depends strongly on the anion, ranging from heavily distorted (RbCeO$_2$: $\theta = 98.6$) to nearly isotropic (RbCeTe$_2$: $\theta = 92.8$). Similarly, the anion plays a critical role in the Ce-Ce distance and the resulting Ce-Ce interactions.}
\label{fig:crystal}
\end{figure}

\begin{figure*}
\centering
\includegraphics[width=7.05in]{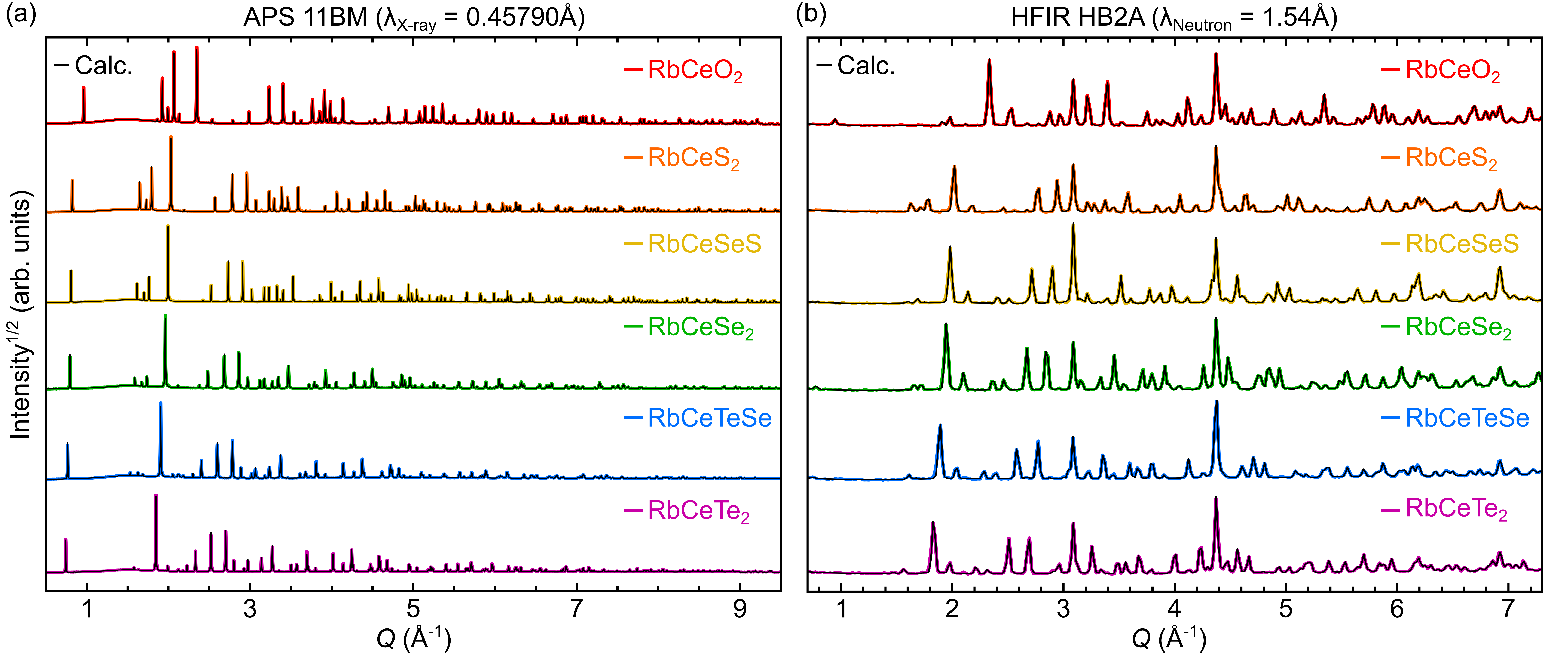}
\caption{(a) Synchrotron x-ray diffraction measurements demonstrate high crystallinity and purity of the RbCeX$_2$ family of compounds. There are trace amounts of nonmagnetic CeO$_2$ in the oxide, and trace amounts of nonmagnetic Rb$_2$Te in the two tellurium-containing compounds. (b) Neutron diffraction measurements corroborate x-ray results and provide more substantial constraints on thermal parameters and, in particular, allow more definitive structural solution of the new oxide, RbCeO$_2$. Fits shown in both (a,b) are results of co-refinement of x-ray and neutron results and include any trace impurities ($<2$\%), kapton background, and aluminum can contributions within the refinements.}
\label{fig:scattering}
\end{figure*}

The CEF Hamiltonian was diagonalized in the CEF interface of {\sc mantid}\cite{arnold2014mantid} to determine energy eigenvalues and eigenvectors of the $J = 5/2$ excited states. Intramultiplet transition probabilities and g-tensor components were calculated from the resulting wave functions. These values were then fit to integrated intensity ratios of the two excitations in the INS data and the g-factor components from quantum chemical calculations by following the minimization procedure described previously \cite{bordelon2021frustrated}. Initial guesses for the CEF parameters were based upon those obtained for KCeO$_2$ with an analogous numerical error minimization processes used to approach a global minimum for the CEF fits \cite{bordelon2020spin}.

\subsection{\textit{Ab initio} wave function calculations}
To obtain a correct starting picture of the Ce$^{3+}$ 4$f^1$ multiplet structures, {\it ab initio} quantum chemical calculations were performed using the {\sc molpro} \cite{Molpro} program package.
For  this  purpose, a finite cluster composed of a CeX$_6$ (X = O, S, Se, Te) octahedron, the six adjacent Ce ions, and twelve Rb nearest neighbors was employed in the calculations. To describe the crystalline environment, we used a large array of point charges that reproduces the crystalline Madelung field within the cluster region. To create this distribution of point charges we used the {\sc ewald} package \cite{Klintenberg_et_al, Derenzo_et_al}. The quantum chemical investigation was initiated as complete active space self-consistent field (CASSCF) calculations \cite{olsen_bible}; an active space consisting of seven 4$f$ orbitals of the central Ce atom was utilized. The Ce 4$f$ and outermost valence $p$ electrons of the central-octahedron ligands were correlated in multireference configuration-interaction (MRCI) calculations with single and double excitations (MRSDCI) \cite{olsen_bible}. Spin-orbit coupling was accounted for according to the procedure described in \cite{Berning_et_al}. Ce-ion $g$ factors were also computed, following the procedure discussed in Ref.~\cite{Bogdanov_et_al}.

Quasirelativistic pseudopotentials \cite{Dolg_Stoll_Preuss} were employed plus valence basis set of quadruple-$\zeta$ quality for the central Ce ion \cite{Cao_Dolg}, all-electron basis sets of triple-$\zeta$ quality for ligands of the central CeX$_6$ octahedron \cite{Dunning_O,Dunning_S,Dunning_Se,Bross2013}, large-core pseudopotentials that also incorporate the 4$f$ electron for the six Ce nearest neighbors \cite{Dolg1989}, and total-ion potentials for the adjacent twelve Rb species \cite{Fuentealba_1983,Szentpaly_et_al}.

\section{Experimental and Computational Results}

\subsection{Synthesis and Structural Properties}

\begin{figure}
\includegraphics[width=\linewidth]{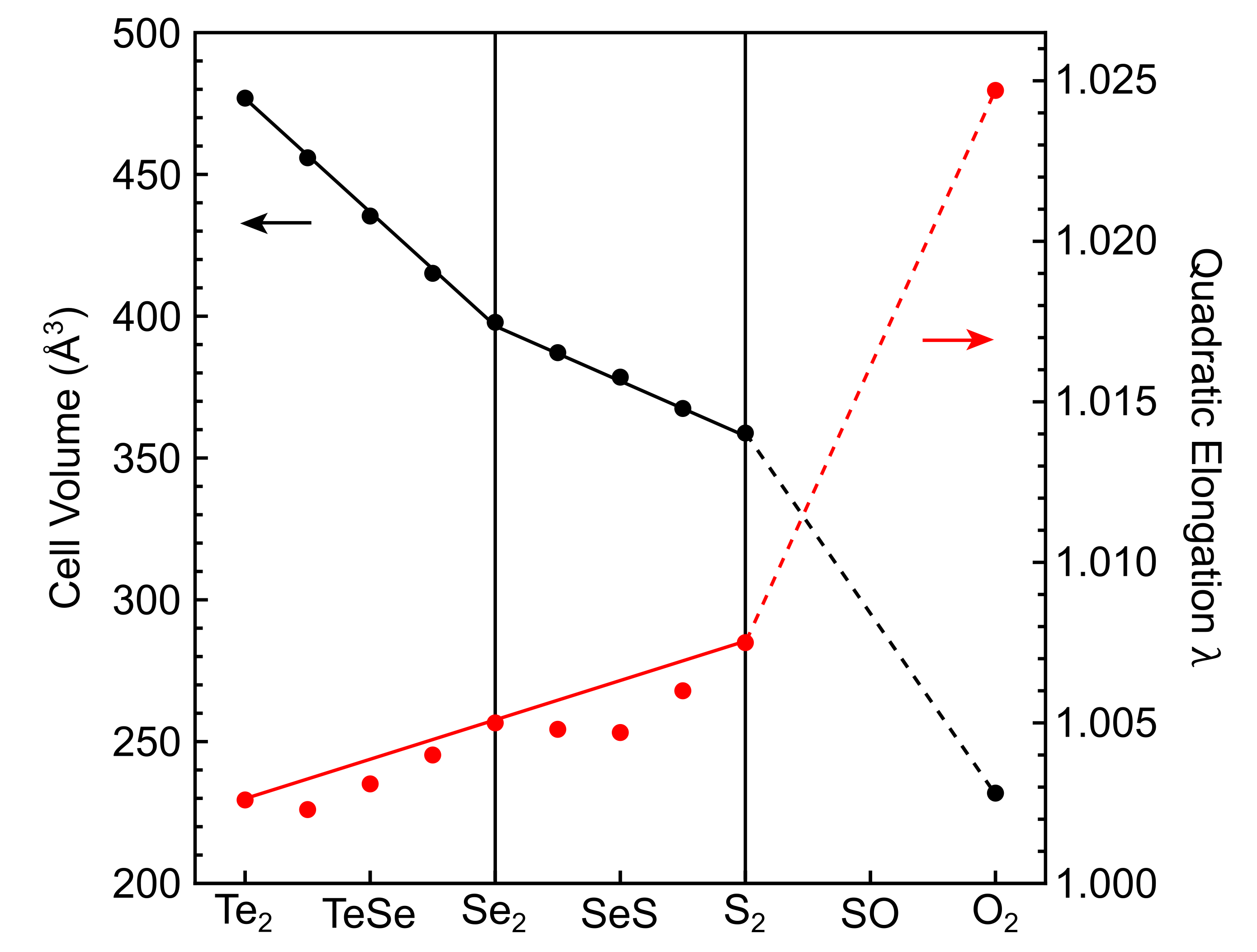}
\caption{Summary of the unit cell volume and the quadratic elongation $\lambda$ for the RbCeX$_2$ alloys. A solubility gap emerges between RbCeS$_2$ and RbCeO$_2$, presumably due to the strong volumetric contraction between the sulfide and oxide. Linear trends between RbCeTe$_2$--RbCeSe$_2$--RbCeS$_2$ confirm a continuous solid solution. The quadratic elongation $\lambda$ provides a dimensionless metric quantifying the distortion of the CeO$_6$ octahedra. While $\lambda$ increases linearly from RbCeTe$_2$ to RbCeS$_2$, intermediate alloys appear to exhibit more complex behavior.}
\label{fig:vegard}
\end{figure}

\begin{table}[h!]
\caption{Crystallographic parameters for the new oxide RbCeO$_2$ and RbCeX$_2$ family, corefined using X-ray synchrotron and neutron data. Anion sites in mixed alloys were constrained to have equivalent adps and z-coordinates. Due to symmetry constraints, $u_{11}$=$u_{22}$=2$u_{12}$ and $u_{13}$=$u_{23}$=$0$.}

	\begin{tabular}{ccccc}
		& & RbCeO$_2$  & & 
		\end{tabular}
    \renewcommand{\tabcolsep}{5.5pt}
	\begin{tabular}{ccccc}
		\hline \hline
		$a$ (\AA) & $c$ (\AA)  & $V$ (\AA$^3$)&  $c/a$ & $\theta$ (\degree) \\ \hline
		3.69562(13) & 19.60128(78) & 231.841(19) & 5.30 & 98.4 \\
		\end{tabular}
	\renewcommand{\tabcolsep}{6pt}
	\begin{tabular}{c|ccccc}
		\hline 
		site & $x$ & $y$ & $z$ & $u_{12}$ & $u_{33}$ \\ \hline
		Rb & 0 & 0 & 0.5           & 0.01034(78)   & 0.00675(26) \\ \hline
		Ce & 0 & 0 & 0             & 0.00813(109)  & 0.00289(5) \\ \hline
		O  & 0 & 0 & 0.27289(5)    & 0.01093(122)   & 0.00557(29) \\ \hline \hline
		\end{tabular}
		
		\hskip+0.3cm
		
	\begin{tabular}{ccccc}
		& & RbCeS$_2$  & & 
		\end{tabular}
    \renewcommand{\tabcolsep}{5.5pt}
	\begin{tabular}{ccccc}
		\hline \hline
		$a$ (\AA) & $c$ (\AA)  & $V$ (\AA$^3$)&  $c/a$ & $\theta$ (\degree) \\ \hline
		4.25219(18) & 22.91427(107) & 358.808(35) & 5.39 & 94.8 \\
		\end{tabular}
	\renewcommand{\tabcolsep}{6.5pt}
	\begin{tabular}{c|ccccc}
		\hline 
		site & $x$ & $y$ & $z$ & $u_{12}$ & $u_{33}$ \\ \hline
		Rb & 0 & 0 & 0.5           & 0.01329(21)   & 0.00875(6) \\ \hline
		Ce & 0 & 0 & 0             & 0.01135(115)  & 0.00301(4) \\ \hline
		S  & 0 & 0 & 0.26693(8)    & 0.00685(143)   & 0.00405(36) \\ \hline \hline
		\end{tabular}

	\hskip+0.3cm

	\begin{tabular}{ccccc}
		& & RbCeSeS  & & 
		\end{tabular}
    \renewcommand{\tabcolsep}{5.5pt}
	\begin{tabular}{ccccc}
		\hline \hline
		$a$ (\AA) & $c$ (\AA)  & $V$ (\AA$^3$)&  $c/a$ & $\theta$ (\degree) \\ \hline
		4.323344(67) & 23.38290(66) & 378.502(16) & 5.41 & 93.8 \\
		\end{tabular}
	\renewcommand{\tabcolsep}{6.5pt}
	\begin{tabular}{c|ccccc}
		\hline 
		site & $x$ & $y$ & $z$ & $u_{12}$ & $u_{33}$ \\ \hline
		Rb & 0 & 0 & 0.5           & 0.01476(16)   & 0.01106(6) \\ \hline
		Ce & 0 & 0 & 0             & 0.01388(10)  & 0.00396(3) \\ \hline
		Se/S  & 0 & 0 & 0.26524(1)    & 0.00727(6)   & 0.00939(22) \\ \hline  \hline
		\end{tabular}

	\hskip+0.3cm

	\begin{tabular}{ccccc}
		& & RbCeSe$_2$  & & 
		\end{tabular}
    \renewcommand{\tabcolsep}{5.5pt}
	\begin{tabular}{ccccc}
		\hline \hline
		$a$ (\AA) & $c$ (\AA)  & $V$ (\AA$^3$)&  $c/a$ & $\theta$ (\degree) \\ \hline
		4.394854(53) & 23.78560(58) & 397.864(14) & 5.41 & 94.0 \\
		\end{tabular}
	\renewcommand{\tabcolsep}{6.7pt}
	\begin{tabular}{c|ccccc}
		\hline 
		site & $x$ & $y$ & $z$ & $u_{12}$ & $u_{33}$ \\ \hline
		Rb & 0 & 0 & 0.5           & 0.01487(24)   & 0.00948(8) \\ \hline
		Ce & 0 & 0 & 0             & 0.01178(19)  & 0.00355(6) \\ \hline
		Se  & 0 & 0 & 0.26561(1)    & 0.00965(40)   & 0.00396(16) \\ \hline \hline
		\end{tabular}

	\hskip+0.3cm

	\begin{tabular}{ccccc}
		& & RbCeTeSe  & & 
		\end{tabular}
    \renewcommand{\tabcolsep}{5.5pt}
	\begin{tabular}{ccccc}
		\hline \hline
		$a$ (\AA) & $c$ (\AA)  & $V$ (\AA$^3$)&  $c/a$ & $\theta$ (\degree) \\ \hline
		4.517862(74) & 24.63007(85) & 435.374(21) & 5.45 & 93.1 \\
		\end{tabular}
	\renewcommand{\tabcolsep}{5.5pt}
	\begin{tabular}{c|ccccc}
		\hline 
		site & $x$ & $y$ & $z$ & $u_{12}$ & $u_{33}$ \\ \hline
		Rb & 0 & 0 & 0.5           & 0.02874(109)   & 0.01260(39) \\ \hline
		Ce & 0 & 0 & 0             & 0.01995(131)  & 0.00456(43) \\ \hline
		Te/Se  & 0 & 0 & 0.26450(1)    & 0.01110(7)   & 0.01575(23) \\ \hline \hline
		\end{tabular}

	\hskip+0.3cm

	\begin{tabular}{ccccc}
		& & RbCeTe$_2$  & & 
		\end{tabular}
    \renewcommand{\tabcolsep}{5.5pt}
	\begin{tabular}{ccccc}
		\hline \hline
		$a$ (\AA) & $c$ (\AA)  & $V$ (\AA$^3$)&  $c/a$ & $\theta$ (\degree) \\ \hline
		4.656079(73) & 25.40110(98) & 476.896(24) & 5.46 & 92.8 \\
		\end{tabular}
	\renewcommand{\tabcolsep}{6.5pt}
	\begin{tabular}{c|ccccc}
		\hline 
		site & $x$ & $y$ & $z$ & $u_{12}$ & $u_{33}$ \\ \hline
		Rb & 0 & 0 & 0.5           & 0.01696(27)   & 0.01111(10) \\ \hline
		Ce & 0 & 0 & 0             & 0.00978(122)  & 0.00522(7) \\ \hline
		Te  & 0 & 0 & 0.26402(1)    & 0.00843(90)   & 0.00559(34) \\ \hline \hline
		\end{tabular}
\label{tab:refinement}
\vspace{-15pt}
\end{table}

To characterize the crystal structure of the newly discovered RbCeO$_2$ and the RbCeX$_2$ (X: S, Se, Te) alloys, x-ray and neutron scattering measurements were performed with the results plotted in Figure \ref{fig:scattering}). Both neutron and synchrotron x-ray results were corefined to provide an comprehensive investigation into the thermal parameters, site occupancies, and crystallography. No significant deviations in stoichiometry or site-mixing were observed within the resolution of the scattering measurements. As such, for the remainder of the analysis in this paper, all atomic occupancies were fixed to 1.0 (fully occupied). A brief summary of the relevant crystallographic parameters for the ternary end members are contained in Table \ref{tab:refinement}.

\begin{figure*}
\centering
\includegraphics[width=7.05in]{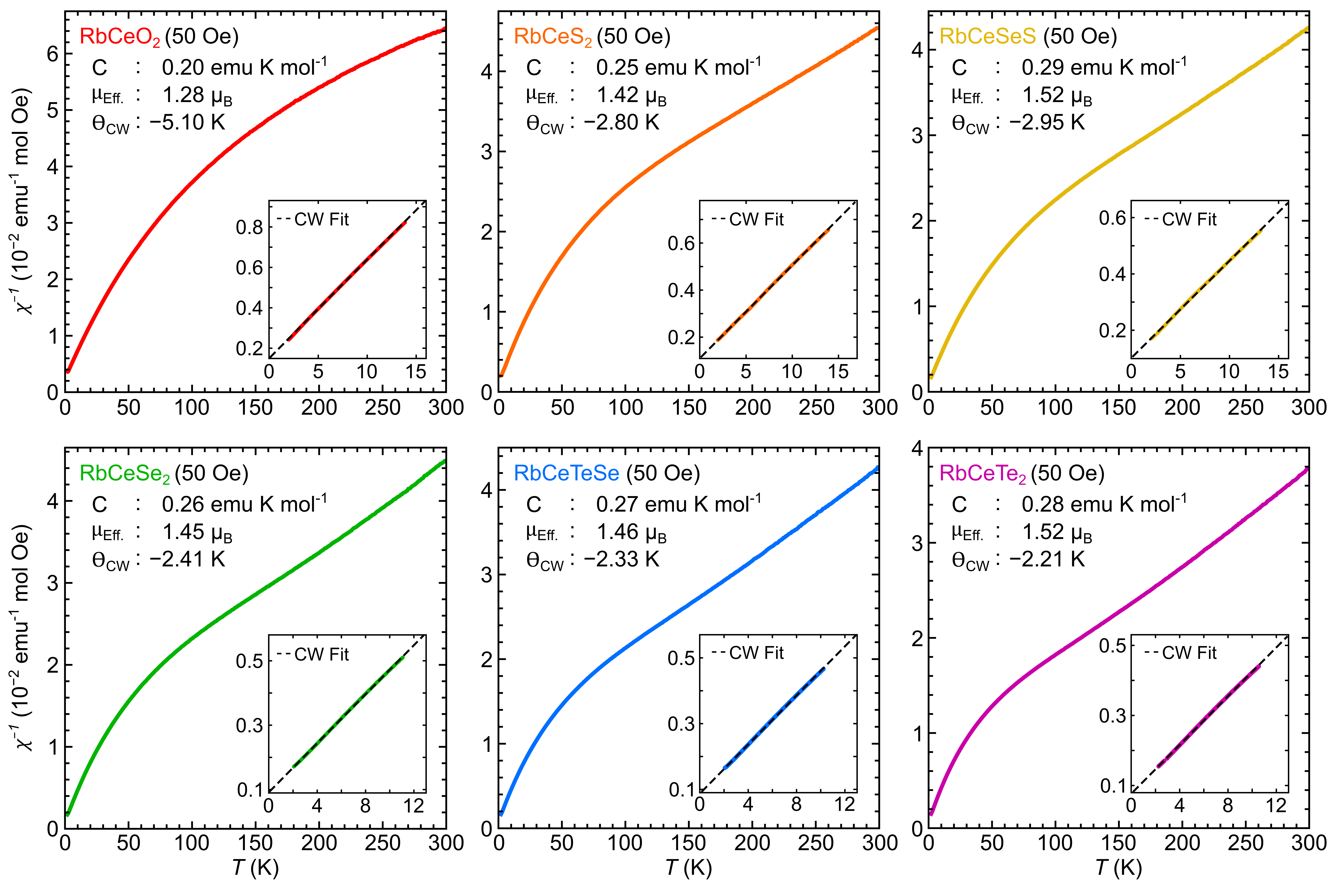}
\caption{Magnetic susceptibility measurements on RbCeX$_2$ series of compounds exhibit no signs of long-range magnetic ordering down to 1.8\,K. Each plot contains an inset that highlights the data range used during the Curie-Weiss analysis. In all cases, strong deviation from linear behavior is noted in the intermediate temperature range ($>50$\,K), likely a contribution from the crystal field levels. The resulting fits (black, dashed) and the fit parameters $\Theta_\text{CW}$, $C$, and $\mu_\text{eff}$ are shown on the individual plots. All compounds exhibit antiferromagnetic coupling between magnetic moments with a effective moment approximately 60-70\% of the expected Ce$^{3+}$ free ion moment.}
\label{fig:magnetism}
\end{figure*}

In addition to the stoichiometric RbCe$X_2$ end members, binary combinations of RbCeSe$_{2-x}$S$_{x}$ and RbCeTe$_{2-x}$Se$_{x}$ were also explored in increments of $x=0.50$ to fully traverse the solid solution. A summary of the cell volume as a function of chemistry is shown in Figure \ref{fig:vegard} for alloys traversing from RbCeTe$_2$ to RbCeO$_2$. The piece-wise linear trend between each ternary end-member indicates a full solid solution extending from RbCeTe$_2$ to RbCeS$_2$. Alloys of RbCeTe$_{2-x}$S$_{x}$ were also attempted; however sample crystallinity was persistently poor and additional peaks were present. At this time it is not clear whether the additional peaks are a result of an impurity, or whether it indicates structural symmetry breaking of the host lattice, as impurity peaks could not be reliably indexed. 

Alloys of the sulfoxide RbCeS$_{2-x}$O$_x$ also failed to produce single-phase $R\overline{3}m$ structures due to phase competition with Ce$_2$O$_2$S.  A strong volumetric contraction is observed upon moving from RbCeS$_2$ to RbCeO$_2$. The internal strain resulting from mixing sulfur and oxygen could intuitively explain the lack of a RbCeS$_{2-x}$O$_{x}$ sulfoxide. The total volumetric contraction between RbCeTe$_2$ and RbCeS$_2$ is of similar magnitude, possibly explaining the difficulty in realizing  RbCeTe$_{2-x}$S$_{x}$ alloys.

\begin{figure*}
\centering
\includegraphics[width=7.05in]{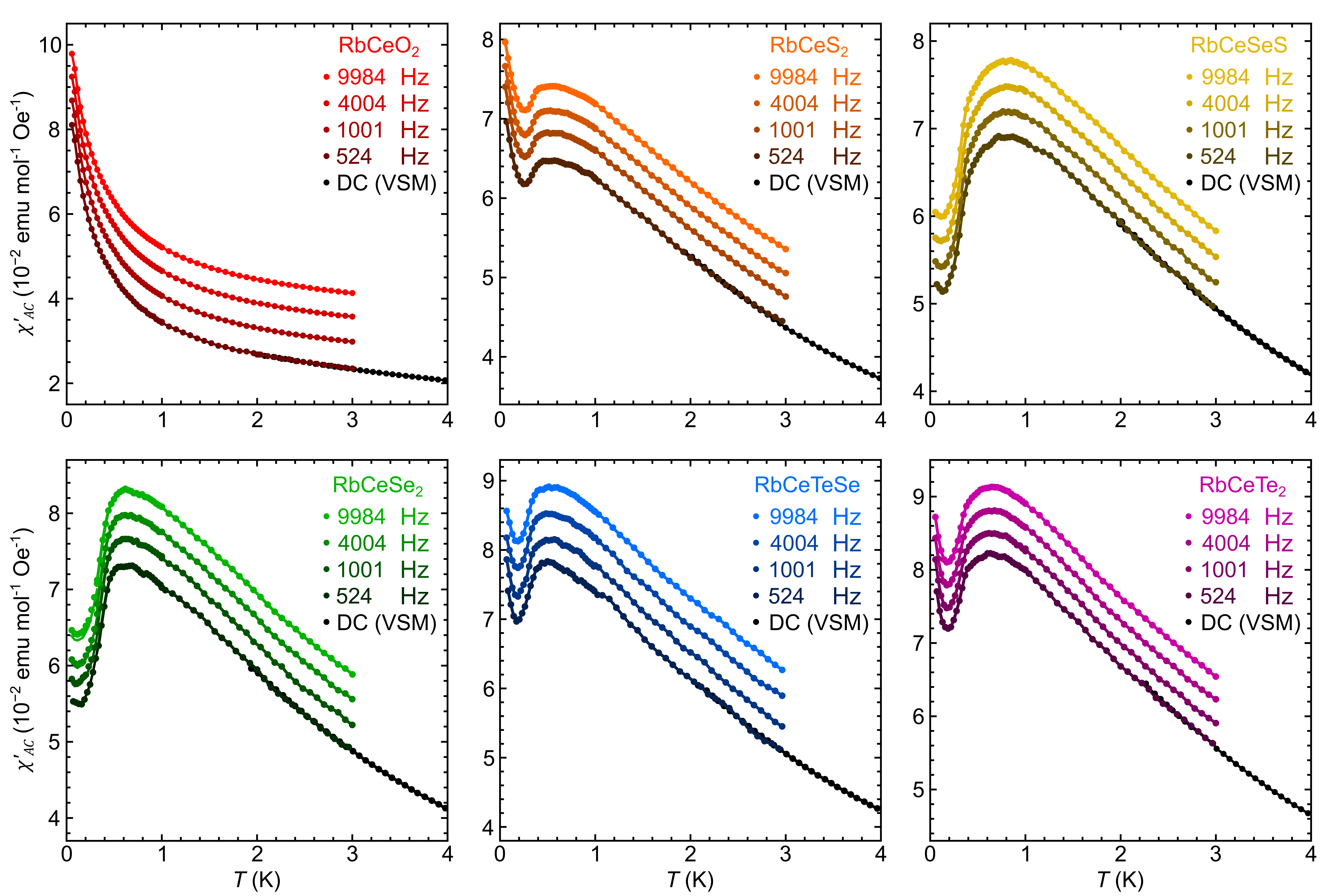}
\caption{Magnetic susceptibility measurements made on RbCeX$_2$ series at temperatures down to 60\,mK. For each compound, the lowest frequency data was scaled to the DC (VSM) magnetization data (black) shown in Figure \ref{fig:magnetism}. The broad peak and inflection in the heavier chalcogenide samples is frequency \textit{independent}, so higher frequency traces are offset vertically for visual clarity in main plots. RbCeO$_2$ resembles transport behavior seen in NaYbO$_2$, exhibiting no signatures of long-range order or freezing. Remarkably, RbCeO$_2$ exhibits no ordering down to 60\,mK, resembling the behavior seen in NaYbO$_2$.}
\label{fig:magnetismDR}
\end{figure*}

Also shown in Figure \ref{fig:vegard} is the quadratic elongation parameter $\lambda$ \cite{robinson1971quadratic}. This parameter provides a dimensionless metric for the distortion of the CeO$_6$ octahedra within the series. Again, we note a striking difference between RbCeO$_2$ and the rest of the RbCe$X_2$ series. RbCeO$_2$ exhibits strongly sheared octahedra with angles that deviate nearly 8.6\degree\, from 90\degree\, (compared with a 2.8\degree\, deviation in RbCeTe$_2$). Considering only the stoichiometric ternary endpoints, the quadratic elongation also trends in a linear fashion until the solubility gap between RbCeS$_2$ and RbCeO$_2$ is reached. However, the intermediate alloys do not trend linearly with the ternary endpoints and instead exhibit local minima within the intermediate compositions.

\subsection{Magnetic Properties} 

To explore the magnetic properties of the RbCeX$_2$ family of compounds, magnetic susceptibility and isothermal magnetization measurements were performed on the ternary compounds RbCeO$_2$, RbCeS$_2$, RbCeSe$_2$, and RbCeTe$_2$ as well as the quaternary alloys RbCeTeSe and RbCeSeS. Susceptibility data collected down to 1.8~K, are shown in Figure \ref{fig:magnetism}. Data were collected under a 50\,Oe DC field in both field-cooled (FC) and zero field-cooled (ZFC) conditions, though no substantial deviations are observed between the FC and ZFC measurements in any sample down to 1.8\,K. 

All compounds within the RbCe$X_2$ series show substantial deviation from Curie-Weiss-like behavior in the moderate to high temperature regime ($>$50\,K). The onset of non-linearity roughly tracks inversely with the anion size, with RbCeO$_2$ showing the most extended linear regime and RbCeTe$_2$ showing the least. As shown later, this arises from the low-lying crystal field levels in the heavy chalcogenide compounds moving downward in energy. For our Curie-Weiss analysis, we first identified a low-temperature window where the thermal population of the first excited state doublet could be avoided. This region, and the corresponding Curie-Weiss fit, is shown as an inset for each RbCe$X_2$ compound.

Upon performing Curie-Weiss fits in this low temperature regime, all compounds show antiferromagnetic Weiss fields $\Theta_\text{CW}$. RbCeO$_2$ exhibits the strongest coupling with $\Theta_\text{CW} = -5.10$\,K, which is comparable to prior results on KCeO$_2$. As with the crystallographic parameters, the properties of the oxide deviate from the rest of the RbCe$X_2$ ($X$: S, Se, Te) series. A substantial drop in $\Theta_\text{CW}$ occurs between RbCeO$_2$ and RbCe$_2$, and continues to decrease in magnitude across the series. RbCeTe$_2$ exhibits the weakest interactions in the series, with $\Theta_\text{CW} = -2.21$\,K. 

Due to the weak interaction strength between Ce-moments, no signatures of long-range magnetic order appear in the magnetization data down to 1.8\,K. To explore the magnetic susceptibility at lower temperatures, AC-susceptibility measurements were performed down to 60 mK at a number of frequencies as shown in Figure \ref{fig:magnetismDR}. The lowest frequency ACDR data (524\,Hz) data were scaled to the DC (VSM) susceptibility data over the temperature range from 2-3\,K. Notably, the oxide shows a low-temperature susceptibility distinct from the other RbCeX$_2$ compounds. All non-oxide compounds (including the alloys) exhibit a peak and subsequent downturn below 0.5--1\,K. RbCeO$_2$ exhibits low-temperature susceptibility measurements remarkably similar to NaYbO$_2$.\cite{nyo_bordelon2019field} Both materials exhibit no signatures of moment freezing, frequency-dependence, short-range, or long-range order down to 60~mK. A more in-depth comparison between the RbCeX$_2$ series and other $ARX_2$ delafossite compounds is provided in the discussion section.

\begin{figure}
\includegraphics[width=\linewidth]{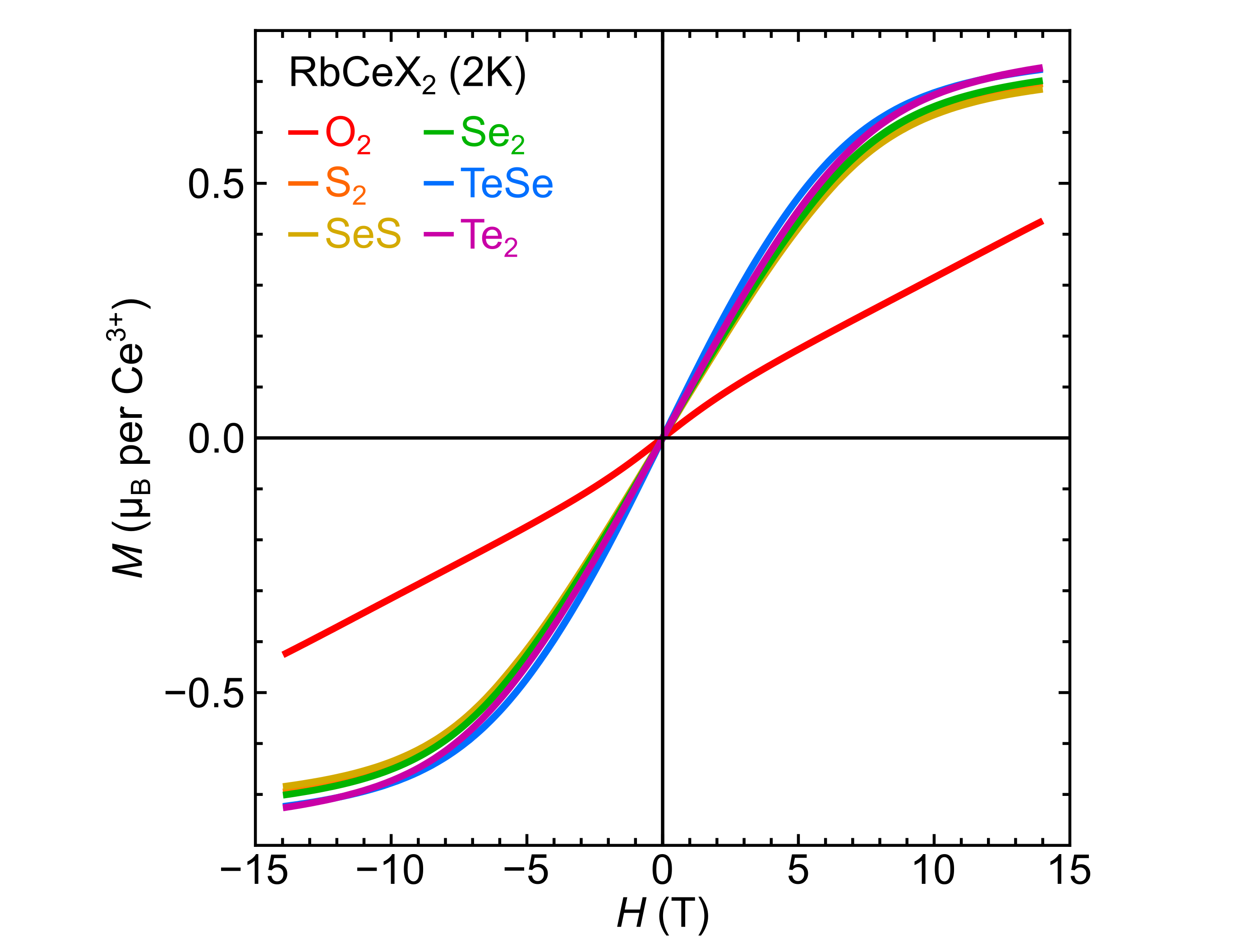}
\caption{Isothermal magnetization results at 2\,K indicate that all heavy chalcogenide RbCeX$_2$ compounds approach the expected saturation magnetization of 0.87$\mu_\text{B}$ by 14\,T. However, the magnetization for RbCeO$_2$ exhibits nearly linear dependence on the field, with no clear indication of saturation at these fields. Rough estimation from the Curie-Weiss fits places the saturation magnetization of RbCeO$_2$ at 0.71$\mu_\text{B}$.}
\label{fig:mvh}
\end{figure}

The differences between the oxide and the other RbCe$X_2$ compounds are also resolved in isothermal magnetization data. Magnetization data collected at 2\,K and plotted in Figure \ref{fig:mvh} indicate that all compounds other than the oxide approach saturation by 14\,T. This is reflective of the substantially larger exchange field in RbCeO$_2$ resolved in Curie-Weiss fits, and the estimated saturated moment for RbCeX$_2$ (X: S, Se, Te) is $g_j J = 0.88\mu_\text{B}$, while the oxide is projected to saturate at $g_j J = 0.74\mu_\text{B}$. 
These estimates are derived from Curie-Weiss fits from Figure \ref{fig:magnetism}, where the expected ``effective'' moment $\mu_\text{eff}$ ($\mu_\text{eff} = \sqrt{8C}$) and the powder averaged $g$-factor $g_\text{avg} = \mu_\text{eff}/\sqrt{J_\text{eff}(J_\text{eff}+1)}$ where $J_\text{eff} \approx 1/2$. 

\subsection{Crystalline Electric Field Analysis}

\begin{figure*}[t]
\includegraphics[width=7.05in]{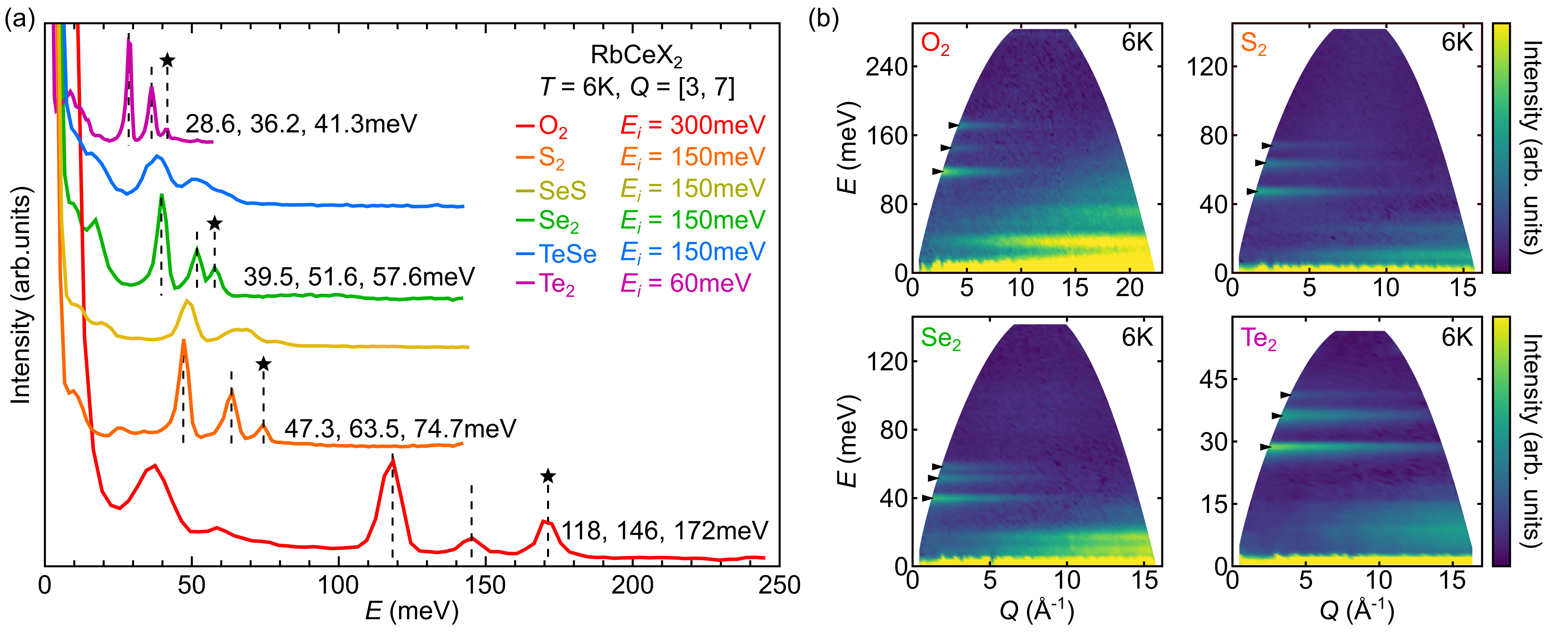}
\caption{(a) Integration of inelastic spectra from Q=[3,7] as a function of energy highlights the individual modes. The energies of each mode are identified on the graphic, extra modes ($E_3$) not predicted by theory are indicated with black stars. Note that the alloyed (RbCeTeSe, RbCeSeS) compounds also exhibit three modes, although the modes are substantially broadened. (b) Background subtracted inelastic spectra for RbCeX$_2$ (X: O, S, Se, Te) series exhibit three clear features consistent with crystalline electric field modes.  }
\label{fig:CEF}
\end{figure*}

Inelastic neutron scattering data exploring the local $J=5/2$ intramultplet excitations for the RbCeX$_2$ (X: O, S, Se, Te) series are presented in Figure \ref{fig:CEF}(a). To optimize data collection for each energy range, a variety of $E_i$'s were used. With the exception of RbCeTe$_2$ (60\,meV) and RbCeO$_2$ (300\,meV), an $E_i$ of 150\,meV was sufficient to capture all of the the CEF modes with sufficient resolution and intensity. 


In all RbCeX$_2$ materials we observe three well-defined, local, modes that decrease in intensity with $Q$ and possess the appropriate energy scale for CEF transitions. Figure \ref{fig:CEF} shows $Q$-averaged line cuts ($Q$=[3,7\AA$^{-1}$]) through I($Q$,E) plotted as a function of energy. The spectral weights of these peaks were used to analyze the CEF level structure. As was previously reported in KCeO$_2$ \cite{kco_bordelon2021magnetic}, there is one extra mode in the experimental spectra that is not normally expected for trivalent Ce$^{3+}$ in the $D_\text{3d}$ symmetry. Recalling that Ce$^{3+}$ (4$f^1$, $L$=3, $S$=1/2) possesses total angular momentum $J$=$|L-S|$=5/2, the $J$=5/2 manifold is ideally split into three CEF doublets according to Kramer's theorem. This suggests that, at maximum, only two intramultiplet excitations should be observed in the INS spectra. Intermultiplet excitations to higher $J$-states ($\approx 250$ meV) are known to only occur at energies far above the window explored here \cite{kco_bordelon2021magnetic}. 

The presence of an anomalous extra mode poses a challenge, as it obscures which modes arise from the conventional CEF-driven splitting of the $J=5/2$ multiplet. To address this, quantum chemical calculations were employed to guide analysis. Experimental CIF files generated from the co-refinement of x-ray and neutron diffraction data were used as a basis for computations within the MOLPRO quantum chemical package. This approach has demonstrated success in modeling the ground states and CEF spectra of Ce-based delafossites \cite{bhattacharyya2022crystal,eldeeb2020energy}.  Relativistic, correlated calculations were performed for (CeX$_6$)Ce$_6$Rb$_{12}$ clusters ($X$: O, S, Se, Te) in order to identify the CEF levels for the 4$f^1$ ion. Relative to the ground state, the first and second excited CEF modes are calculated to be: $E_1$=114.9, $E_2=$146.1\,meV for RbCeO$_2$, $E_1=$47.7, $E_2=$63.4\,meV for RbCeS$_2$,  $E_1=$40.8, $E_2=$55.4\,meV for RbCeSe$_2$, and $E_1=$27.5, $E_2=$34.1\,meV for RbCeTe$_2$. Full tables of the energy landscape for the Ce$^{3+}$ ion in the RbCe$X_2$ family can be found in the supplementary information \cite{ESI}.

The calculations suggest that the highest energy mode in each spectrum ($E_3$) is the anomalous, extra mode. In each case, excitations to the third spin-orbit excited state cannot explain the close proximity of the anomalous mode. For the following analysis of the CEF Hamiltonian using neutron scattering data, the spectral weight of this extra mode is omitted, and its potential origin is further analyzed in the discussion section of this paper.  Using the chosen $E_1$ and $E_2$ excitations, the INS data can be used to fit a model CEF Hamiltonian, where for $D_\text{3d}$ symmetry and Ce$^{3+}$ ($J$=5/2) ions, the Hamiltonian can be expressed in terms of the CEF parameters $B_n^m$ and the Steven's operators $\hat{O}_m^n$:

\begin{equation}
        H_\text{CEF} = B_2^0 \hat{O}_2^0 + B_4^0 \hat{O}_4^0 + B_4^3 \hat{O}_4^3
\end{equation}

\renewcommand{\tabcolsep}{5.8pt}
\begin{table*}
	\caption{Fit and ground state CEF wave functions for RbCeX$_2$ compounds minimized with observed (Obs.) parameters extracted from $E_i$ = 300, 150, and 60 meV INS data. Corresponding Stevens parameters $B_n^m$ and anisotropic $g$ tensor components are shown for each fit.}
	\begin{tabular*}{17.0cm}{l|ccc|ccc|c||ccc|cc}
		\hline
		&&Obs.&&&Fit&&&&&&&  \\ 
		& $E_1$   & $E_2$ & $\frac{I_2}{I_1}$ & $E_1$ & $E_2$ & $\frac{I_2}{I_1}$  & $\chi^2$ & $B_2^0$ & $B_4^0$  & $B_4^4$ & $g_\perp$ & $g_\parallel$   \\ \hline	
		RbCeO$_2$  & 117.7 & 145.2 & 0.286 & 117.8 & 145.2 & 0.392 & 0.03924 & 5.2724 & -0.22591 & -5.08763    & 2.246 & 0.207      \\ \hline 
		
		RbCeS$_2$   & 47.3 & 63.5  & 0.527 & 47.3 & 63.7 & 0.535 & 0.00103 & 1.9995 & -0.07946 & -2.58138  & 2.108 & 0.069   \\  \hline
		
		RbCeSe$_2$   & 39.6 & 51.6  & 0.416 & 39.6 & 51.6 & 0.415 & 0.00010 & 1.8575 & -0.07150 & -1.85899  & 2.225 & 0.164  \\  \hline
		
		RbCeTe$_2$   & 28.7 & 36.2  & 0.655 & 28.6 & 36.3 & 0.663 & 0.00121 & 0.9105 & -0.04862 & -1.62071  & 1.969 & 0.347  \\  \hline \hline
	\end{tabular*}
	\hskip+0.1cm
	\begin{tabular*}{12cm}{l|l}
		Fit wave functions: RbCeO$_2$ & Fit wave functions: RbCeS$_2$ \\
		$| \omega_{0, \pm}\rangle =$  $\pm0.935|\pm1/2\rangle + 0.356|\mp5/2\rangle$ & $| \omega_{0, \pm}\rangle =$  $\mp0.905|\pm1/2\rangle - 0.424|\mp5/2\rangle$ \\		
		$| \omega_{1, \pm}\rangle =$  $1|\pm3/2\rangle$ & $| \omega_{1, \pm}\rangle =$  $1|\pm3/2\rangle$ \\
		$| \omega_{3, \pm}\rangle =$  $\pm0.356|\mp1/2\rangle  + 0.935|\pm5/2\rangle$ & $| \omega_{3, \pm}\rangle =$  $0.424|\mp1/2\rangle  \pm 0.905|\pm5/2\rangle$\\ 
	\end{tabular*}
	\hskip+0.1cm
	\begin{tabular*}{12cm}{l|l}
		\hline
		Fit wave functions: RbCeSe$_2$ & Fit wave functions: RbCeTe$_2$ \\
		$| \omega_{0, \pm}\rangle =$  $\pm0.930|\pm1/2\rangle + 0.367|\mp5/2\rangle$ & $| \omega_{0, \pm}\rangle =$  $\pm0.875|\pm1/2\rangle + 0.484|\mp5/2\rangle$\\		
		$| \omega_{1, \pm}\rangle =$  $1|\pm3/2\rangle$ & $| \omega_{1, \pm}\rangle =$  $1|\pm3/2\rangle$\\
		$| \omega_{3, \pm}\rangle =$  $\pm0.367|\mp1/2\rangle  + 0.930|\pm5/2\rangle$ & $| \omega_{3, \pm}\rangle =$  $\pm0.484|\mp1/2\rangle  + 0.875|\pm5/2\rangle$\\ 
	\end{tabular*}
	\label{tab:tabCEF}
\end{table*}

The CEF Hamiltonian was diagonalized to extract energy eigenvalues and eigenvectors of the $J$=5/2 excited states. Intramultiplet transition probabilities and $g$ tensor components were calculated from the resulting wave functions and were fit to the integrated intensity ratios from the INS data and the $g$-factor results from the quantum chemical calculations. The latter indicate strong $g$-factor anisotropy in all chalcogenides, with MRCI values $g_\text{ab}$=1.46, $g_\text{c}$=0.01 in the RbCeO$_2$, $g_\text{ab}$=1.78, $g_\text{c}$=0.38 in RbCeS$_2$,
$g_\text{ab}$=1.86, $g_\text{c}$=0.32 in RbCeSe$_2$, and $g_\text{ab}$=1.68, $g_\text{c}$=0.81 in RbCeTe$_2$. 

We stress here that the presence of an extra mode and the isolation of two ``intrinsic" modes is an unconventional approach as it ignores portions of the INS spectral weight.  In order to render the effective level scheme that arises from this analysis to be self consistent, the quantum chemistry derived g-factors were used for determining $g_{avg}$ in fitting the neutron scattering data (rather than the experimentally derived $g_{avg}$ from Curie-Weiss fits).  The differences between these values is at most 20\% and the choice of $g_{avg}$ does not qualitatively change the nature of the level-scheme determined.

The results for RbCe$X_2$ (X: O, S, Se, Te) are reported in Table \ref{tab:tabCEF}.  In prior work on KCeO$_2$, the experimental $I_2/I_1$ ratio was significantly overestimated by the fit (0.639 fit, 0.257 exp), suggesting that some of the spectral weight was distributed into the $E_3$ mode. Here we see that the oxide is again overestimated by the fit (0.392 fit, 0.286 exp), although the discrepancy is substantially smaller than in KCeO$_2$. Furthermore, the $I_2/I_1$ ratio for the rest of the series is within expectations from the CEF fit. 

\section{Discussion}

The evolution of the structural and magnetic properties across the RbCe$X_2$ series indicates that the large distortion of the CeO$_6$ octahedra in the oxide variant has a profound effect on the magnetic ground state. The mean-field antiferromagnetic coupling is nearly doubled relative to the other compounds in the series and the apparent local moment is diminished, suggestive of stronger quantum fluctuations in the oxide. The low-temperature susceptibility of RbCeO$_2$ is remarkably similar to that reported in NaYbO$_2$,\cite{nyo_bordelon2019field} where an intrinsic quantum disordered ground state manifests.  While empirically the persistence of the Curie-like feature in RbCeO$_2$ and the absence of a cusp denoting short-range correlations or moment freezing is analogous to that of NaYbO$_2$, further characterization is required to unambiguously determine the ground state properties of the new oxide RbCeO$_2$.

The dramatic shift in properties between RbCeO$_2$ and the remainder of the RbCeX$_2$ compounds is striking. All non-oxide compounds exhibit broad features in their susceptibility at $\sim 0.6-0.8$~K. 
These features can be reasonably ascribed to either: 1) the onset of long-range order or 2) a build-up of short-range magnetic correlations, which seems to vary across the $A$Ce$X_2$ series. 

For $X$=Se, neighboring K- and Cs-based variants have been studied previously, and crystals of KCeSe$_2$ exhibit a similarly broad feature in susceptibility measurements around 0.8~K. Heat capacity measurements on this same system, however, do not observe a sharp inflection, suggesting of a lack of long-range magnetic order and instead only the build up of short-range correlations \cite{sanjeewa2022synthesis}. Crystals of CsCeSe$_2$ exhibit similar properties, showing only a broad feature in $\chi'$ centered at approximately 0.9~K \cite{xing2019crystal}.  For stronger ligand fields, $X$=S and O, the related compounds KCeS$_2$ and KCeO$_2$ instead form long-range order below sharp anomalies in their low temperature heat capacity. KCeS$_2$ and KCeO$_2$ both show transitions at 0.38\,K \cite{bastien2020long} and 0.3~K \cite{kco_bordelon2021magnetic} respectively. Given this ambiguity between long and short-range correlations from susceptibility data alone, the peaks in low-$T$ susceptibility in RbCe(S,Se,Te)$_2$ should be taken as the onset of magnetic correlations and future heat capacity studies would help constrain the state further. The onset of these correlations is notably absent in the oxide RbCeO$_2$.




Turning to the CEF spectra and the effect of sterics, the stronger CEF that arises from having shorter Ce-ligand bonds and stronger distortion of the anions about Ce$^{3+}$ in the oxide strongly affects level splittings. The relative energy of the first excited doublet is increased from 47 meV to 118 meV upon transitioning from the sulfide to the oxide; however both of these energy scales are substantially larger than the magnetic exchange.  Instead, the larger distortion parameter of the CeO$_6$ octahedra and the substantially closer Ce-Ce distances ($3.695$~\AA~ for RbCeO$_2$ versus $4.252$~\AA~  for RbCeS$_2$) generate a much stronger frustrated magnetic exchange field.  These more energetic fluctuations are the likely driver of the onset of potential quantum disorder in RbCeO$_2$ relative to the rest of the RbCe$X_2$ series, though variations in the degree of $g$-factor anisotropy may also play a role.

The presence of an extra mode in the CEF spectrum of RbCe$X_2$ is an interesting and seemingly generic property of $A$Ce$X_2$ delafossites. It was previously noted in the study of KCeO$_2$, where previous calculations and analysis of common defects in KCeO$_2$ have shown that trivial defect states cannot account for the third anomalous CEF mode \cite{kco_bordelon2021magnetic}. More generally, similar modes have been identified in other Ce$^{3+}$ $D_{3d}$ containing compounds as well, including KCeS$_2$ \cite{bastien2020long}, Ce$_2$Zr$_2$O$_7$ \cite{CeZrO_gao2019experimental,CeZrO_gaudet2019quantum}, and Ce$_2$Sn$_2$O$_7$ \cite{CeSnO_sibille2015candidate}. Anomalous CEF modes for lanthanide ions in similar environments have been explained by invoking phonon/vibronic effects \cite{vib_adroja2012vibron,vib_babkevich2015neutron,vib_ellens1997spectral,vib_loong1999dynamic,vib_thalmeier1982bound,vib_thalmeier1984theory,vib_vcermak2019magnetoelastic}, hydrogen impregnation \cite{hyd_rush1980neutron,hyd_wirth1997hydrogen,CeZrO_gao2019experimental,CeZrO_gaudet2019quantum,CeSnO_sibille2015candidate,bastien2020long}, or changes in the local environment due to chemical impurities or disorder \cite{imp_li2017crystalline,bastien2020long,imp_gaudet2015neutron,imp_gaudet2018effect}.  Vibronic effects, such as a dynamic Janh-Teller excitation, are potentially consistent with the systematics of the RbCe$X_2$ series; however the precise mechanism responsible for the anomalous mode's formation is currently undetermined.   

Local chemical perturbations due to the CEF levels are illustrated by the influence of 50\% anion site disorder in RbCeSeS or RbCeTeSe, where the entire spectrum broadens substantially. The presence of three modes remains visible at a relative intensity not unlike the pure ternary systems, which argues against any fine-tuned resonance between local states. Instead the $E_3$ mode tracks downward with increasing $X$-site oscillator mass, consistent with a hybrid mode mixing CEF states with vibrations of the Ce$X_6$ octahedra.  It is notable that the $E_3$ mode has a resolution-limited lifetime, implying an unusually long-lived hybrid state.  This effect seems \textit{endemic} to the Ce$^{3+}$ $D_{3d}$ environment in the RbCe$X_2$ compounds, and motivates further modeling efforts into the potential origin.

\section{Conclusion}

An investigation into the evolution of the structural and magnetic properties across the RbCeX$_2$ series of delafossite compounds was presented. Along with the demonstration of full solid-solutions between RbCeS$_2$--RbCeSe$_2$--RbCeTe$_2$, the discovery of a new oxide, RbCeO$_2$ was reported. Magnetic properties measurements demonstrate that the heavy chalcogenide compounds and their alloys all exhibit features consistent with a buildup of magnetic fluctuations followed by the onset of low-temperature order or short-range correlations. RbCeO$_2$, however, possesses an enhanced antiferromagnetic exchange field and does not exhibit signatures of magnetic ordering down to 60\,mK, suggesting a highly dynamic and disordered ground state. 

Inelastic neutron spectra combined with quantum chemical computations and model-Hamiltonian simulations across the RbCeX$_2$ series further determine the CEF level schemes and ground state wave functions. Consistent with prior work, an additional mode appears in the CEF spectra of all RbCeX$_2$ compounds  -- demonstrating that this mode is \textit{endemic} to the RbCeX$_2$ series and motivates further research into its origin. These results highlight the importance of the anion component in ground state formation in the broader rare-earth delafossite series and further encourage study of rare-earth delafossite compounds as tunable platforms for the exploration of quantum disordered magnetic ground states.

\section{Acknowledgments}

S.D.W. acknowledges fruitful discussions with Andrew Christianson, Stephan Rosenkranz, and Eric Bauer.  This work was supported by the U.S. Department of Energy, Office of Basic Energy Sciences, Division of Materials Sciences and Engineering under award DE-SC0017752 (S.D.W., B.R.O., P.M.S., G.P. and M.B.). Experiments used facilities supported via the UC Santa Barbara NSF Quantum Foundry funded via the Q-AMASE-i program under award DMR-1906325. B.R.O. and P.M.S. also acknowledge support from the California NanoSystems Institute through the Elings fellowship program. P.B., T.P., and L.H. thank U.~Nitzsche for technical assistance and the German Research Foundation (Grant Nos. 441216021 and 437124857) for financial support. Use of the Advanced Photon Source at Argonne National Laboratory was supported by the U. S. Department of Energy, Office of Science, Office of Basic Energy Sciences, under Contract No. DE-AC02-06CH11357. This research used resources at the High Flux Isotope Reactor (beamline HB-2A) and the Spallation Neutron Source (beamline BL-18 ARCS), which are DOE Office of Science User Facilities operated by the Oak Ridge National Laboratory.

\bibliography{RbCeX2_v6.bib}

\end{document}